\date{12 February 2006}

\newif\ifelsevier
\newif\ifarxiv
\arxivtrue

\documentclass[twocolumn,prd]{revtex4}

\ifelsevier

\fi

\usepackage{graphicx}
\usepackage{psfrag}
\usepackage{url}
\usepackage{units}
\usepackage{amsmath}

\newcommand{\MSbar}{\ensuremath{ \overline{\rm MS} }}

\newcommand{\Lag}{\mathcal{L}}

\newcommand{\eqdef}{\stackrel{\textrm{def}}{=}}

\newcommand{\eqbad}{\stackrel{\textrm{?}}{=}}

\DeclareRobustCommand*\diff[2][]{%
   \mathop{\mathrm{{d}^{#1}}{}#2}\nolimits
}

\newcommand{\VC}[1]{%
  \begin{tabular}[c]{l}%
    #1%
  \end{tabular}
}

\begin{document}
\title{Renormalization: general theory%
\ifarxiv 
   \footnote{%
      To appear in the Encyclopedia of Mathematical Physics (Elsevier, 2006).
   }%
\fi
}
\author{John C. Collins\\
        Physics Department,
        Penn State University,\\
        104 Davey Laboratory,
        University Park PA 16802-6300, USA
}

\maketitle

\section{Introduction}

Quantum field theories (QFTs) provide a natural framework for quantum
theories that obey the principles of special relativity.  Among their
most striking features are ultra-violet divergences, which at first
sight invalidate the existence of the theories.  The divergences arise
from Fourier modes of very high wave number, and hence from the
structure of the theories at very short distances.  In the very
restricted class of theories called ``renormalizable'', the
divergences may be removed by a singular redefinition of the
parameters of the theory.  This is the process of renormalization, that
defines a QFT as a non-trivial limit of a theory with a UV cut-off.

A very important QFT is the Standard Model, an accurate and successful
theory for all the known interactions except gravity.  Calculations
using renormalization and related methods are vital to the theory's
success.

The basic idea of renormalization predates QFT.  Suppose we treat an
observed electron as a combination of a bare electron of mass $m_0$
and the associated classical electromagnetic field down to a radius
$a$.  The observed mass of the electron is its bare mass plus the
energy in the field (divided by $c^2$).  The field energy is
substantial, e.g., $0.7\,{\rm MeV}$ when $a=\unit[10^{-15}]{m}$, and
it diverges when $a\to0$.  The observed mass, $0.5\,{\rm MeV}$, is the
sum of the large (or infinite) field contribution compensated by a
negative and large (or infinite) bare mass.  This calculation needs
replacing by a more correct version for short distances, of course,
but it remains a good motivation.

In this article, I review the theory of renormalization in its classic
form, as applied to weak-coupling perturbation theory, or Feynman
graphs.  It is this method, rather than the Wilsonian approach
reviewed elsewhere in this volume, that is typically used in practice
for perturbative calculations in the Standard Model, especially its
QCD part.

Much of the emphasis is on weak-coupling perturbation theory, where
there are well-known algorithmic rules for performing calculations and
renormalization. Applications --- see the article on QCD and
confinement for some important non-trivial examples --- involve
further related results, such as the operator product expansion,
factorization theorems, and the renormalization group, to go far
beyond simple fixed order perturbation theory.  The construction of
fully rigorous mathematical treatments for the exact theory is a topic
of future research.

\section{Formulation of QFT}

A QFT is specified by its Lagrangian density.  A simple example is
$\phi^4$ theory:
\begin{equation}
\label{eq:L.original}
   \Lag
   \eqbad
   \frac{(\partial\phi)^2}{2} -\frac{m^2\phi^2}{2}
   -\frac{\lambda \phi^4}{4!} ,
\end{equation}
where $\phi(x)=\phi(t,\3x)$ is a single component hermitian field.  The
Lagrangian density and the resulting equation of motion,
$\partial^2\phi+m^2\phi+\frac16\lambda\phi^3=0$, are local; they involve only products of
fields at the same space-time point.  Such locality is characteristic
of relativistic theories, where otherwise it is difficult or
impossible to preserve causality, but it is also the source of the UV
divergences.  The question mark over the equality symbol in Eq.\ 
(\ref{eq:L.original}) is a reminder that renormalization of UV
divergences will force us to modify the equation.

The Feynman rules for perturbation theory are given by a free
propagator $i/(p^2-m^2+i0)$ and a interaction vertex $-i\lambda$.  Although
we will usually work in four space-time dimensions, it is useful also
to consider the theory in a general space-time dimensionality $n$,
where the coupling has energy dimension $[\lambda]=E^{4-n}$.  We use
``natural units'', i.e., with $\hbar=c=1$.
The ``$i0$'' in the propagator $i/(p^2-m^2+i0)$ symbolizes the
location of the pole relative to the integration contour; it is often
written as $i\epsilon$.

The primary targets of calculations are the vacuum expectation values
of time-ordered products of $\phi$; in QFT these are called the Green
functions of the theory.  From these can be reconstructed the
scattering matrix, scattering cross sections, and other measurable
quantities.

\section{One-loop calculations}

\begin{figure}
    \centering
       $ \VC{\includegraphics[scale=0.65]{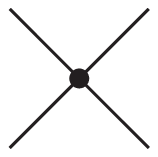}}
        + \VC{\includegraphics[scale=0.65]{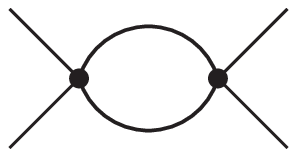}}
        + \VC{\includegraphics[scale=0.65]{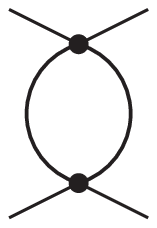}}
        + \VC{\includegraphics[scale=0.65]{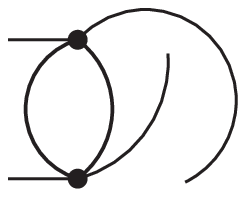}}
        + O(\lambda^3)
       $
    \caption{One-loop approximation to connected and amputated 4-point
      function, before renormalization.}
\label{fig:1loop4}
\end{figure}

Low-order graphs for the connected and amputated 4-point Green
function are shown in Fig.~\ref {fig:1loop4}.  Each one-loop graph has
the form
\begin{align}
\label{eq:1loop}
   -i\lambda^2I(p^2)
   \eqbad&
   \frac { \lambda^2 }{ 2 }
          \int \frac{ \diff[4]{k} }{ (2\pi)^4 }
          \frac{1}
               { (k^2-m^2+i0 ) \, [(p-k)^2-m^2+i0 ] },
\end{align}
where $p$ is a combination of external momenta.  There is a divergence
from where the loop momentum $k$ goes to infinity.  We define the
degree of divergence, $\Delta$, by counting powers of $k$ at large
$k$, to get $\Delta=0$.  In an $n$-dimensional space-time we would have
$\Delta=n-4$.  The integral is divergent whenever $\Delta\geq0$.  Comparing the
dimensions of the one-loop and tree graphs shows that $\Delta$ equals the
negative of the energy-dimension of the coupling $\lambda$.  Thus the
dimensionlessness of $\lambda$ at the physical space-time dimension is
equivalent to the integral being just divergent.

The infinity in the integral implies that the theory \emph{in its
  naive formulation} is not defined.  With the aid of renormalization
group methods, it has been shown that the problem is with the complete
theory, not just perturbation theory.

The divergence only arises because we use a continuum space-time.  So
suppose that we formulate the theory initially on a lattice of spacing
$a$ (in space or space-time). Our loop graph is now
\begin{equation}
   -i\lambda^2I(p;m,a) 
   =
   \frac{ -\lambda^2 }{32\pi^4} \int \diff[4]{k}
   S(k,m;a) \, S(p-k,m;a),
\end{equation}
where the free propagator $S(k,m;a)$ approaches the usual value
$i/(k^2-m^2+i0)$ when $k$ is much smaller than $1/a$, and it falls off
more rapidly for large $k$.  The basic observation that propels the
renormalization program is that the divergence as $a\to0$ is independent
of $p$.  This is most easily seen by differentiating once with respect
to $p$, after which the integral is convergent when $a=0$, because the
differentiated integral has degree of divergence $-1$.

\begin{figure}
    \centering
       $ \VC{\includegraphics[scale=0.60]{figures/4pt-LO}}
        \psfrag{A}{$3A$}
        + \VC{\includegraphics[scale=0.60]{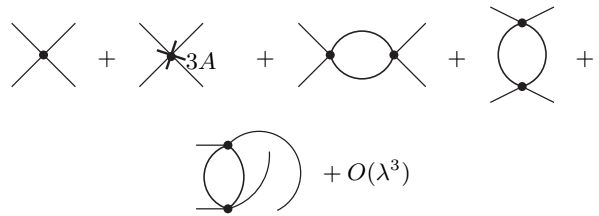}}
        + \VC{\includegraphics[scale=0.60]{figures/4pt-NLO-s}}
        + \VC{\includegraphics[scale=0.60]{figures/4pt-NLO-t}}
        + \VC{\includegraphics[scale=0.60]{figures/4pt-NLO-u}}
        + O(\lambda^3)
       $
    \caption{One-loop approximation to renormalized connected and
      amputated 4-point function, with counterterm.}
\label{fig:1loop4.ren}
\end{figure}

Thus we can cancel the divergence in Eq.\ (\ref{eq:1loop}) by
replacing the coupling in the first term in Fig.~\ref{fig:1loop4}, by
the so-called bare coupling
\begin{equation}
\label{eq:bare.1}
  \lambda_0 = \lambda + 3A(a)\lambda^2 + O(\lambda^3).
\end{equation}
Here $A(a)$ is chosen so that the renormalized value of our one-loop
graph,
\begin{equation}
\label{eq:ren.1}
  -i\lambda^2I_R(p^2,m^2)
  =
   -i\lambda^2 \lim_{a\to0}
        \left[ I(p;m,a) + A(a) \right],
\end{equation}
exists, at $a=0$, with $A(a)$ in fact being real-valued.  The factor
$3$ multiplying $A(a)$ in Eq.\ (\ref{eq:bare.1}) is because there are
three one-loop graphs, with equal divergent parts.  The replacement
for the coupling is made in the tree graph in Fig.~\ref {fig:1loop4},
but not yet at the vertices of the other graphs, because at the moment
we are only doing a calculation accurate to order $\lambda^2$; the
appropriate expansion parameter of the theory is the finite
renormalized coupling $\lambda$, held fixed as $a\to0$.  We call the extra
term in Eq.\ (\ref{eq:ren.1}) a counterterm.  The diagrams for the
correct renormalized calculation are represented in Fig.~\ref
{fig:1loop4.ren}, which has a counterterm graph compared with
Fig.~\ref {fig:1loop4}.

In the physics terminology, used here, the cutting-off of the
divergence by using a modified theory is called a regularization.
This contrasts with the mathematics literature, where ``regularized
integral'' usually means the same as a physicist's ``renormalized
integral''.

There is always freedom to add a finite term to a counterterm.  When
we discuss the renormalization group, we will see that this
corresponds to a reorganization of the perturbation expansion and
provides a powerful tool for improving perturbatively based
calculations, especially in QCD. Contrary to the impression given in
some parts of the literature, it is not necessary that a renormalized
mass equal a corresponding physical particle mass, with similar
statements for coupling and field renormalization.  While
such a prescription is common and natural in a simple theory like QED,
it is by no means required and certainly may not always be best.  If
nothing else, the correspondence between fields and stable particles
may be poor or non-existent (as in QCD).

One classic possibility is to subtract the value of the graph at
$p=0$, a prescription associated with Bogoliubov, Parasiuk and Hepp
(BPH), which leads to
\begin{align}
   -i\lambda^2I_{R,\,{\rm BPH}}(p^{2})
  = & 
    \frac{-i\lambda^2}{32\pi^2}
    \int_0^1 \diff{x} \ln\left[ 1-p^2x(1-x)/m^2 \right] .
\end{align}
In obtaining this from (\ref{eq:1loop}), we used a standard Feynman
parameter formula,
\begin{equation}
    \frac{1}{AB} = \int_0^1 \diff{x} \frac{1}{ [Ax + B(1-x)]^2 },
\end{equation}
to combine the propagator denominators, after which the integral over
the momentum variable $k$ is elementary.  We then
obtain the renormalized one-loop (4-point and amputated) Green
function
\begin{equation}
  -i\lambda -i\lambda^2 \left[ I_R(s) + I_R(t) + I_R(u) \right] + O(\lambda^3) ,
\end{equation}
where $s$, $t$, and $u$ are the three standard Mandelstam invariants
for the Green function.  [For a $2\to2$ scattering process, or a
corresponding off-shell Green function, in which particles of momenta
$p_1$ and $p_2$ scatter to particles of momenta $p_1'$ and $p_2'$, the 
Mandelstam variables are
defined as $s=(p_1+p_2)^2$, $t=(p_1-p_1')^2$, and $u=(p_1-p_2')^2$.]

In the general case, with a nonzero degree of divergence, the
divergent part of an integral is a polynomial in $p$ and $m$ of degree
$D$, where $D$ is the smallest positive integer less than or equal to
$\Delta$.  In a higher space-time dimension, this implies that
renormalization of the original, momentum-independent, interaction
vertex is not sufficient to cancel the divergences.  We would need
higher derivative terms, and this is evidence that the theory is not
renormalizable in higher than 4 space-time dimensions.  Even so, the
terms needed would be local, because of the polynomiality in $p$.


\section{Complete formulation of renormalization program}

The full renormalization program motivated by example calculations is:
\begin{itemize}
\item The theory is regulated to cut off the divergences.
\item The numerical value of each coefficient in $\Lag$ is allowed to
  depend on the regulator parameter (e.g., $a$).
\item These dependences are adjusted so that finite results for Green
  functions are obtained after removal of the regulator.
\end{itemize}
In $\phi^4$ theory, we therefore replace $\Lag$ by
\begin{equation}
\label{eq:L0.1}
   \Lag = \frac{Z}{2}(\partial\phi)^2 - \frac{ Zm_0^2 }{2} \phi^2 
                 - \frac{ Z^2\lambda_0 }{4!} \phi^4,
\end{equation}
with the bare parameters, $Z$, $m_0$ and $\lambda_0$, having a regulator
dependence such that Green functions of $\phi$ are finite at $a=0$. 

The slightly odd labeling of the coefficients in Eq.\ (\ref{eq:L0.1})
arises because observables like cross sections are invariant under a
redefinition of the field by a factor.  In terms of the bare field
$\phi_0\eqdef\sqrt{Z}\phi$, we have
\begin{equation}
\label{eq:L0.2}
   \Lag = \frac{1}{2} (\partial\phi_0)^2 - \frac{m_0^2}{2} \phi_0^2 
                 - \frac{\lambda_0}{4!} \phi_0^4.
\end{equation}
The unit coefficient of $\frac12(\partial\phi_0)^2$ implies that $\phi_0$ has
canonical commutation relations (in the regulated theory).  This
provides a natural standard for the normalization of the bare
mass $m_0$ and the bare coupling $\lambda_0$.

All terms in $\Lag$ have coefficients with dimension zero or larger.
This is commonly characterized by saying that the terms $\Lag$ ``have
dimension 4 or less'', which refers to the products of field operators
and derivatives in each term.  A generalization of the power-counting
analysis shows that if we start with a theory whose $\Lag$ only has
terms of dimension 4 or less, then no terms of higher dimension are
needed as counterterms, at least not in perturbation theory.  This is
a very powerful restriction on self-contained QFTs, and was critical
in the discovery of the Standard Model.

Sometimes it is found that the description of some piece of physics
appears to need higher dimension operators, as was the case originally
with weak interaction physics.  The lack of renormalizability of such
theories indicates that they cannot be complete, and an upper bound on
the scale of their applicability can be computed, e.g., a few hundred
GeV for the four-fermion theory of weak interactions.  Eventually this
theory was superseded by the renormalizable Weinberg-Salam theory of
weak interactions, now a part of the Standard Model, to which the
four-fermion theory provides a low-energy approximation for charged
current weak interactions.

Certain operators of allowed dimensions are missing in Eq.\ 
(\ref{eq:L0.1}): the unit operator, and $\phi$ and $\phi^3$.  Symmetry under
the transformation $\phi\to-\phi$ implies that Green functions with an odd
number of fields vanish, so that no $\phi$ and $\phi^3$ counterterms are
needed.  Divergences with the unit operator do appear, but not for
ordinary Green functions.  In gravitational physics, the coefficient
of the unit operator gives renormalization of the cosmological
constant.

To implement renormalized perturbation theory, we partition $\Lag$
(non-uniquely) as
\begin{equation}
  \Lag = \Lag_{\rm free} + \Lag_{\rm basic~interaction}
         +\Lag_{\rm counterterm},
\end{equation}
where the free, the basic interaction and the counterterm Lagrangians
are
\begin{align}
\label{eq:L.free}
   \Lag_{\rm free} ={}&
   \frac {1}{2}(\partial\phi)^2 - \frac{m^2}{2} \phi^2 ,
\\
\label{eq:L.basic_int}
   \Lag_{\rm basic~interaction} ={}&
    - \frac {\lambda}{4!} \phi^4 ,
\\
\label{eq:L.ct}
   \Lag_{\rm counterterm} ={}&
    \frac{Z-1}{2} (\partial\phi)^2 
    - \, \frac{ (Zm_0^2-m^2) }{2} \, \phi^2
\nonumber\\&
    - \, \frac{ (Z^2\lambda_0-\lambda) }{4!} \, \phi^4 .
\end{align}
The renormalized coupling and mass, $\lambda$ and $m$, are to be fixed and
finite when the UV regulator is removed.  Both the basic interaction
and the counterterms are treated as interactions.  First we compute
``basic graphs'' for Green functions using only the basic interaction.
The counterterms are expanded in powers of $\lambda$, and then all graphs
involving counterterm vertices at the chosen order in $\lambda$ are added to
the calculation.  The counterterms are arranged to cancel all the
divergences, so that the UV regulator can be removed, with $m$ and $\lambda$
held fixed.  The counterterms cancel the parts of the basic Feynman
graphs associated with large loop momenta.  An algorithmic
specification of the otherwise arbitrary finite parts of the
counterterms is called a renormalization prescription or a
renormalization scheme.  Thus it gives a definite relation between the
renormalized and bare parameters, and hence a definite specification
of the partitioning of $\Lag$ into its three parts.

It has been proved that this procedure works to all orders in $\lambda$,
with corresponding results for other theories.  Even in the absence
of fully rigorous non-perturbative proofs, it appears clear that the
results extend beyond perturbation theory, at least in asymptotically
free theories like QCD: see the discussion of the Wilsonian RG
elsewhere in this volume.

\section{Dimensional regularization and minimal subtraction}

The final result for renormalized graphs does not depend on the
particular regularization procedure. A particularly convenient
procedure, especially in QCD, is dimensional regularization, where
divergences are removed by going to a low space-time dimension $n$.
To make a useful regularization method, $n$ is treated as a continuous
variable, $n=4-2\epsilon$.

Great advantages of the method are that it preserves Poincar{\'e}
invariance and many other symmetries (including the gauge symmetry of
QCD), and that Feynman graph calculations are minimally more
complicated than for finite graphs at $n=4$, particularly when all the
lines are massless, as in many QCD calculations.  

Although there is no such object as a genuine vector space of finite
non-integer dimension, it is possible to construct an operation that
behaves as if it were an integration over such a space.  The operation
was proved unique by Wilson, and explicit constructions have been
made, so that consistency is assured at the level of all Feynman
graphs.  Whether a satisfactory definition beyond perturbation theory
exists remains to be determined.

It is convenient to arrange that the renormalized coupling is
dimensionless in the regulated theory.  This is done by changing the
normalization of $\lambda$ with the aid of an extra parameter, the unit of
mass $\mu$:
\begin{equation}
  \lambda_0 = \mu^{2\epsilon} \left( \lambda + \text{counterterms} \right),
\end{equation}
with $\lambda$ and $\mu$ being held fixed when $\epsilon\longrightarrow0$.  [Thus the basic
interaction in Eq.\ (\ref{eq:L.basic_int}) is changed to
$-\lambda\mu^{2\epsilon}\phi^4/4!$.]
Then for the one-loop graph of Eq.\ (\ref{eq:1loop}), dimensionally
regularized Feynman parameter methods give
\begin{align}
 -i\lambda^2I(p;m,\epsilon) 
 ={}&
 \frac{i\lambda^2}{32\pi^2} (4\pi)^\epsilon \Gamma(\epsilon)
\nonumber\\
& 
    \int_0^1 \diff{x}
        \left[
           \frac{ m^2-p^2x(1-x) -i0 }{ \mu^2 }
        \right]^{-\epsilon} .
\end{align}
A natural renormalization procedure is to subtract the pole at $\epsilon=0$,
but it is convenient to accompany this with other factors to remove
some universally occurring finite terms.  So \MSbar{} renormalization
(``modified minimal subtraction'') is defined by using the counterterm
\begin{equation}
\label{eq:MSbar.def}
 -iA(\epsilon)\lambda^2 = -i \frac{\lambda^2S_\epsilon}{32\pi^2\epsilon} ,
\end{equation}
where $S_\epsilon \eqdef (4\pi e^{-\gamma_E})^\epsilon$, with $\gamma_E = 0.5772\dots$ being the
Euler constant.  This gives a renormalized integral (at $\epsilon=0$)
\begin{equation}
\label{eq:MSbar.ex1}
 -\frac{i\lambda^2}{32\pi^2} 
    \int_0^1 \diff{x}
        \ln\,\left[
           \frac{ m^2-p^2x(1-x) }{ \mu^2 }
        \right],
\end{equation}
which can be evaluated easily.  A particularly simple result is
obtained at $m=0$:
\begin{equation}
\label{eq:MSbar.ex1a}
 \frac{i\lambda^2}{32\pi^2} 
        \left[
          -\ln \frac{ -p^2}{ \mu^2 }
          +2
        \right].  
\end{equation}
This formula symptomizes important and very useful algorithmic
simplifications in the higher-order massless calculations common in
QCD.

The \MSbar{} scheme amounts to a \textit{de facto} standard for QCD.
At higher orders a factor of ${S_\epsilon}^L$ is used in the counterterms, with
$L$ being the number of loops.

\section{Coordinate-space}

Quantum fields are written as if they are functions of $x$, but they
are in fact distributions or generalized functions, with
quantum-mechanical-operator values.  This indicates that using
products of fields is dangerous and in need of careful definition.
The relation with ordinary distribution theory is simplest in the
coordinate-space version of Feynman graphs.  Indeed in the 1950s,
Bogoliubov and Shirkov formulated renormalization as a problem of
defining products of the singular numeric-valued distributions in
coordinate-space Feynman graphs; theirs was perhaps the best
treatment of renormalization in that era.

For example, the coordinate-space version of Eq.\ (\ref{eq:ren.1}) is
\begin{align}
\label{eq:ren.coord.1}
&
   -\lambda^2 \lim_{a\to0}
        \int \diff[4]{x} \diff[4]{y}
        f(x,y)
\nonumber\\ &\hspace*{1cm}
        \left[ \frac12 \tilde{S}(x-y;m,a)^2 + iA(a)\delta^{(4)}(x-y) \right],
\end{align}
where $x$ and $y$ are the coordinates for the interaction vertices,
$f(x,y)$ is the product of external-line free propagators, and
$\tilde{S}(x-y;m,a)$ is the coordinate-space free propagator, which at
$a=0$ has a singularity
\begin{equation}
  \frac{ 1 }{ 4\pi^2 \, [-(x-y)^2+i0] }
\end{equation}
as $(x-y)^2\to0$.  We see in Eq.\ (\ref{eq:ren.coord.1}) a version of
the Hadamard finite-part of a divergent integral, and renormalization
theory generalizes this to particular kinds of arbitrarily
high-dimension integrals.  The physical realization and
\emph{justification} of the use of the finite-part procedure is in
terms of renormalization of parameters in the Lagrangian; this also
gives the procedure a significance that goes beyond the integrals
themselves and involves the full non-perturbative formulation of QFT.

\section{General counterterm formulation}

We have written $\Lag$ as a basic Lagrangian density plus
counterterms, and have seen in an example how to cancel divergences at
one-loop order.  In this section, we will see how the procedure works
to all orders.  The central mathematical tool is Bogoliubov's
$R$-operation.  Here the counterterms are expanded as a sum of terms,
one for each basic 1PI graph with a non-negative degree of divergence.
To each basic graph for a Green function is added a set of counterterm
graphs associated with divergences for subgraphs. The central theorem
of renormalization is that this procedure does in fact remove all the
UV divergences, with the form of the counterterms being determined by
the simple computation of the degree of divergence for 1PI graphs.

\begin{figure}
    \begin{center}
    \leavevmode
        $\VC{\includegraphics[scale=0.65]{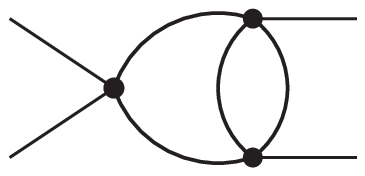}}
        \psfrag{A}{$2A$}
        ~+~ \VC{\includegraphics[scale=0.65]{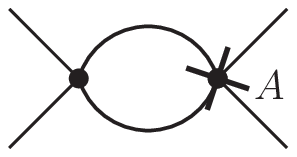}}
        \psfrag{A}{$B$}
        ~+~ \VC{\includegraphics[scale=0.65]{figures/4pt-LO-ct}}
        $
    \end{center}
    \caption{A 2-loop graph and its counterterms. The label $B$
      indicates that it is the two-loop overall counterterm for this
      graph. }
\label{fig:2loop}
\end{figure}

To see the essential difficulty to be solved, consider a two-loop
graph like the first one in Fig.\ \ref{fig:2loop}.  Its divergence is
not a polynomial in external momenta, and is therefore not canceled by
an allowed counterterm.  This is shown by differentiation with respect
to external momenta, which does not produce a finite result because of
the divergent one-loop subgraph.  But for consistency of the theory,
the one-loop counterterms already computed must be themselves put into
loop graphs.  Among others, this gives the second graph of Fig.\ 
\ref{fig:2loop}, where the cross denotes that a counterterm
contribution is used.  The contribution used here is actually $2/3$ of
the total one-loop counterterm, for reasons of symmetry factors that
are not fully evident at first sight.  The remainder of the one-loop
coupling renormalization cancels a subdivergence in another 2-loop
graph.  It is readily shown that the divergence of the sum of the
first two graphs in Fig.\ \ref{fig:2loop} is momentum-independent, and
thus can be canceled by a vertex counterterm.

This method is fully general, and is formalized in the Bogoliubov
$R$-operation, which gives a recursive specification of the
renormalized value $R(G)$ of a graph $G$:
\begin{equation}
   R(G) \eqdef G + \sum_{ \{\gamma_1,\dots,\gamma_n\} } 
                \left. G \right|_{\gamma_i\to C(\gamma_i)} .
\label{eq:RDef}
\end{equation}
The sum is over all sets of non-intersecting one-particle-irreducible
(1PI) subgraphs of $G$, and the notation $G|_{\gamma_i\to C(\gamma_i)}$ denotes
$G$ with all the subgraphs $\gamma_i$ replaced by associated counterterms
$C(\gamma_i)$.  The counterterm $C(\gamma)$ of a 1PI graph $\gamma$ has the form
\begin{align}
   C(\gamma) \eqdef{}& - T \left(
                  \gamma  + \mbox{Counterterms for subdivergences}
              \right) .
\label{eq:CDef}
\end{align}
Here $T$ is an operation that extracts the divergent part of its
argument and whose precise definition gives the renormalization
scheme.  For example, in minimal subtraction we define
\begin{equation}
   T(\Gamma) = \mbox{pole part at $\epsilon=0$ of $\Gamma$}.
\end{equation}
We formalize the term inside parentheses in Eq.\ (\ref{eq:CDef}) as:
\begin{align}
   \bar{R}(\gamma) \eqdef & \gamma + \mbox{Counterterms for subdivergences}
\nonumber\\
    =& \gamma + \mathop{\sum\nolimits'}\limits_{ \{\gamma_1,\dots,\gamma_n\} }
                \left. G \right|_{\gamma_i\to C(\gamma_i)} ,
\label{eq:RPrimeDef}
\end{align}
where the prime on the $\sum\nolimits'$ denotes that we sum over all sets
of non-intersecting 1PI subgraphs except for the case that there is a
single $\gamma_i$ equal to the whole graph (i.e., the term with 
$n=1$ and $\gamma_1=\gamma$ is omitted).

Note that, for the \MSbar{} scheme, we define the $T$ operation to be
applied to a factor of constant dimension obtained by taking the
appropriate power of $\mu^\epsilon$ outside of the pole part operation.
Moreover it is not a strict pole part operation; instead each pole is
to be multiplied by ${S_\epsilon}^L$, where $L$ is the number of loops, and
$S_\epsilon$ is defined after Eq.\ (\ref{eq:MSbar.def}).

Eqs.\ (\ref{eq:RDef})--(\ref{eq:RPrimeDef}) give a recursive
construction of the renormalization of an arbitrary graph.  The
recursion starts on one-loop graphs, since they have no
subdivergences, i.e., $C(\gamma)=-T(\gamma)$ for a one-loop 1PI graph.

Each counterterm $C(\gamma)$ is implemented as a contribution to the
counterterm Lagrangian.  The Feynman rules ensure that once $C(\gamma)$ has
been computed, it appears as a vertex in bigger graphs in such a way
as to give exactly the counterterms for subdivergences used in the
$R$-operation.  It has been proved that the $R$-operation does in fact
give finite results for Feynman graphs, and that basic power-counting
in exactly the same fashion as at one-loop determines the relevant
operators.

\begin{figure}
  \centering
  \includegraphics[scale=0.65]{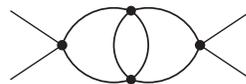}
  \caption{Graph with overlapping divergent subgraphs.}
  \label{fig:overlap}
\end{figure}

In early treatments of renormalization, a problem was caused by graphs
like Fig.\ \ref{fig:overlap}.  This graph has three divergent
subgraphs which overlap, rather than being nested.  Within the
$R$-operation approach, such cases are no harder to deal with than
merely nested divergences.

The recursive specification of $R$-operation can be converted to a
non-recursive formulation by the forest formula of Zavyalov and
Stepanov, later rediscovered by Zimmermann.  It is normally the
recursive formulation that is suited to all-orders proofs. 

Whether these results, proved to all orders of perturbation theory,
genuinely extend to the complete theory is not so easy to answer,
certainly in a realistic 4-dimensional QFT.  One illuminating case is
of a non-relativistic quantum mechanics model with a delta-function
potential in a two-dimen\-sional space.  Renormalization can be applied
just as in field theory, but the model can also be treated exactly and
it has been shown that the results agree with perturbation theory.

Perturbation series in relativistic QFTs can at best be expected to be
asymptotic, not convergent.  So instead of a radius of convergence, we
should talk about a region of applicability of a weak coupling
expansion.  In a direct calculation of counterterms, etc, the radius
of applicability shrinks to zero as the regulator is removed.  However
we can deduce the expansion for a renormalized quantity, whose
expansion is expected to have a nonzero range of applicability.  We
can therefore appeal to the uniqueness of power series expansions to
allow the calculation, at intermediate stages, to use bare quantities
that are divergent as the regulator is removed.

\section{Renormalizability, non-renormalizability and
  super-renormalizability}  

The basic power counting method shows that if a theory with
conventional fields (at $n=4$) has only operators of dimension 4 or
less in its $\Lag$, then the necessary counterterm operators are also
of dimension 4 or less.  So if we start with a Lagrangian with all
possible such operators, given the field content, then the theory is
renormalizable.  This is not the whole story, as we will see in the
discussion of gauge theories.

If we start with a Lagrangian containing operators of dimension higher
than 4, then renormalization requires operators of ever higher
dimension as counterterms when one goes to higher orders in
perturbation theory.  Therefore, such a theory is said to be
perturbatively non-renormalizable.  Some very powerful methods of
cancellation or some non-perturbative effects are needed to evade this
result.

In the case of dimension-4 interactions, there is only a finite set of
operators given the set of basic fields, but divergences occur at
arbitrarily high orders in perturbation theory.  If, instead, all the
operators have at most dimension 3, then only a finite number of
graphs need counterterms.  Such theories are called
superrenormalizable.  The divergent graphs also occur as subgraphs
inside bigger graphs, of course.  There is only one such theory in a
4-dimensional space-time: $\phi^3$ theory, which suffers from an energy
density that is unbounded from below, so it is not physical.  In lower
space-time dimension, where the requirements on operator dimension are
different, there are many more known super-renormalizable theories,
some with a very rigorous proof of existence.

All the above characterizations rely primarily on perturbative
analysis, so they are subject to being not quite accurate in an exact
theory, but they form a guide to the relevant issues.

\section{Renormalization and symmetries; gauge theories}

In most physical applications, we are interested in QFTs whose
Lagrangian is restricted to obey certain symmetry requirements.  Are
these symmetries preserved by renormalization?  That is, is the
Lagrangian with all necessary counterterms still invariant under the
symmetry?

We first discuss non-chiral symmetries; these are symmetries in which
the left-handed and right-handed parts of Dirac fields transform
identically. 

For Poincar{\'e} invariance and simple global internal symmetries, it is
simplest to use a regulator, like dimensional regularization, which
respects the symmetries.  Then it is easily shown that the
symmetries are preserved under renormalization.  This holds even if
the internal symmetries are spontaneously broken [as happens with a
``wrong-sign mass term'', e.g., negative $m^2$ in Eq.\ 
(\ref{eq:L.original})].

The case of local gauge symmetries is harder.  But their preservation
is more important, because gauge theories contain vector fields which,
without a gauge symmetry, generally give unphysical features to the
theory.  For perturbation theory, BRST quantization is usually used,
in which, instead of gauge symmetry, there is a BRST supersymmetry.
This is manifested at the Green function level by Slavnov-Taylor
identities that are more complicated, in general, than the Ward
identities for simple global symmetries and for abelian local
symmetries.

Dimensional regularization preserves these symmetries and the
Slavnov-Taylor identities.  Moreover the $R$-operation still produces
finite results with local counterterms, but cancellations and
relations occur between divergences for different graphs in order to
preserve the symmetry.  A simple example is QED, which has an abelian
U(1) gauge symmetry, and whose gauge-invariant Lagrangian is
\begin{equation}
  \label{eq:QED}
   \Lag =
    -\frac14 \left( \partial_\mu A^{(0)}_\nu - \partial_\nu A^{(0)}_\mu \right)^2
    + \bar\psi_0 \left( i\gamma^\mu\partial_\mu - e_0A^{(0)}_\mu - m_0 \right) \psi_0.
\end{equation}
\emph{At the level of individual divergent 1PI graphs}, we get
counterterms proportional to ${A_\mu}^2$ and to $({A_\mu}^2)^2$, operators
not present in the gauge-invariant Lagrangian.  The Ward identities
and Slavnov-Taylor identities show that these counterterms cancel when
they are summed over all graphs at a given order of renormalized
perturbation theory.  Moreover the renormalization of coupling and the
gauge field are inverse, so that $e_0A^{(0)}_\mu$ equals the
corresponding object with renormalized quantities, $\mu^\epsilon eA_\mu$.
Naturally, sums of contributions to a counterterm in $\Lag$ can only
be quantified with use of a regulator.

In non-abelian theories the gauge-invariance properties are not just
the absence of certain terms in $\Lag$ but quantitative relations
between the coefficients of terms with different numbers of fields.
Even so, the argument with Slavnov-Taylor identities generalizes
appropriately and proves renormalizability of QCD, for example.  But
note that the relation concerning the product of the coupling and the
gauge field does not generally hold; the form of the gauge
transformation is itself renormalized, in a certain sense.

\section{Anomalies}

Chiral symmetries, as in the weak-interaction part of the gauge
symmetry of the Standard Model, are much harder to deal with.  Chiral
symmetries are ones for which the left-handed and right-handed
components of Dirac field transform independently under different
components of the symmetry group, local or global as the case may be.
Sometimes, some or other of the left-handed or right-handed components
may not even be present.  

In general, chiral symmetries are not preserved by regularization, at
least not without some other pathology.  At best one can adjust the
finite parts of counterterms such that in the limit of the removal of
the regulator, the Ward or Slavnov-Taylor identities hold.  But in
general, this cannot be done consistently, and the theory is said to
suffer from an anomaly.  In the case of chiral gauge theories, the
presence of an anomaly prevents the (candidate) theory from being
valid. A dramatic and non-trivial result (Adler-Bardeen theorem and
some non-trivial generalizations) is that if chiral anomalies cancel
at the one-loop level, then they cancel at all orders.

Similar results, but more difficult ones, hold for supersymmetries.

The anomaly cancellation conditions in the Standard Model lead to
constraints that relate the lepton content to the quark content in
each generation.  For example, given the existence of the $b$ quark,
and the $\tau$ and $\nu_\tau$ leptons (of masses around $\unit[4.5]{GeV}$,
$\unit[1.8]{GeV}$, and zero respectively), it was strongly predicted
on the grounds of anomaly cancellation that there must be a $t$ quark
partner of the $b$ to complete the third generation of quark doublets.
This prediction was much later vindicated by the discovery of the much
heavier top quark with $m_t \simeq \unit[175]{GeV}$.

\section{Renormalization schemes}

A precise definition of the counterterms entails a specification of
the renormalization prescription (or scheme), so that the finite parts
of the counterterms are determined.  This apparently induces extra
arbitrariness in the results.  However, in the $\phi^4$ Lagrangian (for
example), there are really only two independent parameters.  (A
scaling of the field does not affect any observables, so here we do
not count $Z$ as a parameter here.)  Thus at fixed regulator parameter $a$
or $\epsilon$, renormalization actually just gives a reparameterization of a
two-parameter collection of theories.  A renormalization prescription
gives the change of variables between bare and renormalized parameters,
a rather singular transformation when the regulator is removed.  If we
have two different prescriptions, we can deduce a transformation
between the renormalized parameters in the two schemes.  The
renormalized mass and coupling $m_1$ and $\lambda_1$ in one scheme can be
obtained as functions of their values $m_2$ and $\lambda_2$ in the other
scheme, with the bare parameters, and hence the physics, being the
same in both schemes.  Since these are renormalized parameters, the
removal of the regulator leaves the transformation well behaved.

Generalization to all renormalizable theories is immediate.

\section{Renormalization group and applications and generalizations}

One part of the choice of renormalization scheme is that of a scale
parameter such as the unit of mass $\mu$ of the \MSbar{} scheme.  The
physical predictions of the theory are invariant if a change of $\mu$ is
accompanied by a suitable change of the renormalized parameters, now
considered as $\mu$-dependent parameters $\lambda(\mu)$ and $m(\mu)$.  These are
called the effective, or running, coupling and mass.  The
transformation of the parameterization of the theory is called a
renormalization-group (RG) transformation.

The bare coupling and mass $\lambda_0$ and $m_0$ are RG-invariant, and this
can be used to obtain equations for the RG-evolution of the effective
parameters from the perturbatively computed counterterms.  For
example, in $\phi^4$ theory, we have (in the renormalized theory after
removal of the regulator)
\begin{equation}
  \frac{ \diff{\lambda} }{ \diff{\ln\mu^2} }
  = \beta(\lambda),
\end{equation}
with $\beta(\lambda) = 3\lambda^2/(16\pi^2) + O(\lambda^3)$.  As exemplified in Eqs.\ 
(\ref{eq:MSbar.ex1}) and (\ref{eq:MSbar.ex1a}), Feynman diagrams
depend logarithmically on $\mu$.  By choosing $\mu$ to be comparable to
the physical external momentum scale, we remove possible large
logarithms in this and higher orders.  Thus, provided that the
effective coupling at this scale is weak, we get an effective
perturbation expansion.

This is a basic technique for exploiting perturbation theory in QCD,
for the strong interactions, where the interactions are not
automatically weak.  In this theory the RG $\beta$ function is negative so
that the coupling decreases to zero as $\mu\to\infty$; this is the asymptotic
freedom of QCD.  

A closely related method is that associated with the Callan-Symanzik
equation, which is a formulation of a Ward identity for anomalously
broken scale invariance.  However, RG methods are the actually-used
ones, normally, even if sometimes a RG equation is incorrectly
labeled as a Callan-Symanzik equation.

The elementary use of the RG is not sufficient for most interesting
processes, which involve a set of widely different scales.  Then more
powerful theorems come into play.  Typical are the factorization
theorems of QCD, reviewed elsewhere.  These express differential cross
sections for certain important reactions as a product of quantities
that involve a single scale:
\begin{equation}
  \label{eq:fact}
  \diff{\sigma} = C\!\left( Q,\mu,\lambda(\mu) \right) 
             \otimes f\!\left( m,\mu,\lambda(\mu) \right) + \mbox{small correction}.
\end{equation}
The product is typically a matrix or a convolution product.  The
factors obey non-trivial RG equations, and these enable different
values of $\mu$ to be used in the different factors.  Predictions arise
because some factors and the kernels of the RG equation are
perturbatively calculable, with a weak effective coupling. Other
factors, such as $f$ in Eq.\ (\ref{eq:fact}), are not perturbative.
These are quantities with names like ``parton distribution
functions'', and they are universal between many different processes.
Thus the non-perturbative functions can be measured in a limited set
of reactions and used to predict cross sections for many other
reactions with the aid of calculations of the perturbative factors.

Ultimately this whole area depends on physical phenomena associated
with renormalization.


\section{Concluding remarks}

The actual ability to remove the divergences in certain QFTs to
produce consistent, finite and non-trivial theories is a quite
dramatic result.  Moreover, associated with the integrals that give
the divergences is behavior of the kind that is analyzed with
renormalization-group methods and generalizations.  So the properties
of QFTs associated with renormalization get tightly coupled to many
interesting consequences of the theories, most notably in QCD.  

Quantum field theories are actually very abstruse and difficult
theories; only certain aspects currently lend themselves to practical
calculations.  So the reader should not assume that all aspects of
their rigorous mathematical treatment are perfect.  Experience, both
within the theories and in their comparison with experiment,
indicates, nevertheless, that we have a good approximation to the
truth. 

When one examines the mathematics associated with the $R$-operation
and its generalizations with factorization theorems, there are clearly
present some interesting mathematical structures that are not yet
formulated in their most general terms.  Some indications of this can
be seen in the work by Connes and Kreimer, reviewed in another
article, where it is seen that renormalization is associated with a
Hopf algebra structure for Feynman graphs.  

With such a deep subject, it is not surprising that it lends itself to
other approaches, notably the Connes-Kreimer one and the Wilsonian
one, also reviewed in another article.  Readers new to the subject
should not be surprised if it is difficult to get a fully unified view
of these different approaches.


\section*{Acknowledgments}

I would like to thank Prof.\ K. Goeke for hospitality and support at
the Ruhr-Universit{\"a}t-Bochum.  This work was also supported by the U.S.
D.O.E.


\section*{See also}

Quantum Chromodynamics.  Exact Renormalization Group.  Perturbative
Renormalization Theory and BRST.  Effective Field Theories.  Quantum
Field Theory: A Brief Introduction. Hopf Algebra Structure of
Renormalizable Quantum Field Theory.  Lattice Gauge Theory.
Perturbation Theory and its Techniques.  Standard Model of Particle
Physics.  BRST Quantization.  Anomalies.  Electroweak Theory.
Operator Product Expansion in Quantum Field Theory.


\section*{Keywords}
\begin{itemize}
\item Renormalization
\item Relativistic quantum field theory
\item Ultra-violet divergences
\item $R$ operation
\item Minimal subtraction
\end{itemize}




\end{document}